\documentclass[numreferences]{kluwer}

\begin{document}  \sloppy


\begin{article}

\begin{opening}

\title{Vanquishing the XCB Question:
The Methodological Discovery of the Last Shortest Single Axiom for the Equivalential Calculus\thanks{
The work of the first and third authors was supported by the Mathematical, Information, and Computational Sciences Division subprogram of the Office of Advanced Scientific Computing Research, U.S. Department of Energy, under Contract W-31-109-Eng-38.}}

\runningtitle{Vanquishing the XCB Question}

\author{Larry \surname{Wos}}
\institute{Mathematics and Computer Science Division, Argonne National Laboratory, Argonne, IL  60439, U.S.A.
E-mail: wos@mcs.anl.gov
}

\author{Dolph \surname{Ulrich}}
\institute{Department of Philosophy,
Purdue University, West Lafayette, IN  47907-1360, U.S.A.
E-mail: dulrich@purdue.edu
}
\author{Branden \surname{Fitelson}}
\institute{Philosophy Department,
San Jose State University,
San Jose, CA  95192, U.S.A.
E-mail: branden@fitelson.org
}

\date{}

\begin{abstract}
With the inclusion of an effective methodology, this article answers in detail a question that, for a quarter of a century, remained open despite intense study by various researchers.
Is the formula $XCB = e(x,e(e(e(x,y),e(z,y)),z))$ a single axiom for the classical equivalential calculus when the rules of inference consist of detachment ({\em modus ponens}) and substitution?
Where the function $e$ represents equivalence, this calculus can be axiomatized quite naturally with the formulas $e(x,x)$, $e(e(x,y),e(y,x))$, and $e(e(x,y),e(e(y,z),e(x,z)))$, which correspond to reflexivity, symmetry, and transitivity, respectively.
(We note that $e(x,x)$ is dependent on the other two axioms.)
Heretofore, thirteen shortest single axioms for classical equivalence of length eleven had been discovered, and  $XCB$ was the only remaining formula of that length whose status was undetermined.  To show that $XCB$ is indeed such a single axiom, we focus on the rule of condensed detachment, a rule that captures detachment together with an appropriately general, but restricted, form of substitution.
The proof we present in this paper consists of twenty-five applications of condensed detachment, completing with the deduction of transitivity followed by a deduction of symmetry.
We also discuss some factors that may explain in part why $XCB$ resisted relinquishing its treasure for so long.
Our approach relied on diverse strategies applied by the automated reasoning program OTTER.
Thus ends the search for shortest single axioms for the equivalential calculus.
\end{abstract}

\end{opening}

\section{History, Significance, and Terminology}
 
With the use of an effective methodology, we answer the final open question concerning possible shortest single axioms for the classical equivalential calculus.
Specifically, when detachment (that is, {\em modus ponens}) and substitution are the inference rules in use, is the formula

\vspace{.1in}
$e(x,e(e(e(x,y),e(z,y)),z))$ \quad\quad   (XCB)
\vspace{.1in}

\noindent
a single axiom for that calculus, or is it too weak?
This question, posed by J. Peterson [9], remained open for a quarter of a century, eluding the efforts of various researchers.
Here we tell the full story of the discovery of the main result reported in our companion article ``XCB, The Last of the Shortest Single Axioms for the Classical Equivalential Calculus'' [16], where we also treat a simpler open question posed by K. Hodgson [2]:  Is $XCB$ a single axiom for classical equivalential calculus in the presence of substitution, detachment, and reverse detachment?
 
For a formula to be a single axiom for a formal system, all theorems of that system must be deducible from the formula.
We make this perhaps obvious observation because, rather than relying on the combination of detachment and substitution, we study $XCB$ here in the presence of but one inference rule, condensed detachment (discussed more fully shortly), a rule that captures detachment coupled with a constrained form of substitution.
To answer Peterson's question, we study the specific question of whether one can rely solely on condensed detachment to deduce from the formula $XCB$ some known basis (axiom system) for the equivalential calculus.
If that could be proved impossible, then $XCB$ would be too weak to serve as a single axiom (see, for example, [5]).
If, on the other hand, such a deduction can be found, then $XCB$ must itself also be such a basis, and the question open for two and one-half decades is answered in the affirmative.
 
We do in fact answer Peterson's question affirmatively by presenting a proof (in Section 3) obtained with invaluable assistance from W. McCune's automated reasoning program OTTER [7].
Reliance on such a program naturally suggests an attack featuring condensed detachment rather than detachment coupled with substitution.
The former rule is (as will be shown) easily implemented in OTTER through the use of hyperresolution, whereas the latter pair of rules is far less attractive because of the lack of effective strategies for choosing from among the myriad instances obtainable with substitution.
In addition, proofs based on condensed detachment are often more elegant.
Indeed, when one is faced with reading a formula (obtained, say, by substitution) thousands of symbols long, one finds the task daunting.
The proof we give in Section 3 would, if presented in terms of detachment and substitution, include two such formulas.
 
The discovery of this proof, deducing from $XCB$ a known 2-basis consisting of formulas that correspond to symmetry and transitivity, marks a victory both for logic and for automated reasoning [14].
Thus ends the search for shortest single axioms for the equivalential calculus.
Indeed, exactly fourteen such axioms exist, and no others can be found.
 
The use of methodology and strategy (discussed in Section 2) shows that automated reasoning played a vital role.
Before we detail the techniques crucial to our success, we briefly review the equivalential calculus and discuss the relevant contributions made by earlier logicians to the study of shortest single axioms for this field.

\subsection{The Equivalential Calculus}
 
Formulas of the classical equivalential calculus are constructed from sentential variables and the two-place function symbol $e$ (for ``equivalence'').
The theorems of this logic are precisely the formulas in which each variable occurs an even number of times---$e(x,x)$, $e(e(x,y)$,$e(y,x))$, $e(e(y,y)$, $e(y,y))$, and the like.
Those theorems in which each variable present occurs exactly twice are said to have the 2-property.
An interesting and useful fact is that every theorem of the calculus either has the 2-property or is a substitution instance of a theorem with the 2-property.
For example, $e(e(y,y),e(y,y))$ does not have the 2-property, but this formula is an instance of $e(x,x)$, which does.
Also known [1] (compare [4]) is the fact that, when condensed detachment is successfully applied to two formulas each of which has the 2-property, the formula that results from that application always has the 2-property.

As one might guess from its name, the equivalential calculus can be axiomatized (see, for example, [13]) with formulas corresponding to reflexivity, symmetry, and transitivity, expressed here as clauses.
\begin{verbatim}
P(e(x,x)).
P(e(e(x,y),e(e(y,z),e(x,z)))).
P(e(e(x,y),e(y,x))).
\end{verbatim}
Unexpectedly, perhaps, reflexivity is provably dependent on the 2-basis consisting of symmetry and transitivity.
 
In the early years, before C. A. Meredith entered the game, equivalential calculus was studied exclusively in terms of two rules of inference, detachment and substitution.
Detachment permits the deduction of $t$ from the two hypotheses (premisses) $e(s,t)$ and $s$.
Thanks to Meredith [8], however, it has become standard practice to study the equivalential calculus instead in terms of the sole inference rule condensed detachment.
Briefly (see [5] for a detailed presentation), this rule takes as premisses two formulas $e(s,t)$ and $r$ (assumed to share no variables in common) and, when $s$ and $r$ are unifiable, permits the deduction of the formula obtained by applying to $t$ the most general substitution unifying $s$ and $r$.
Condensed detachment can be easily implemented in OTTER with the use of hyperresolution and the following clause as its nucleus, where `` $\mid$ '' denotes logical {\bf or} and `` - '' logical {\bf not}.
\begin{verbatim}
-P(e(x,y)) | -P(x) | P(y).
\end{verbatim}
 
From the viewpoint of substitution and detachment, when attempting to apply condensed detachment, one first seeks a most general substitution of terms for variables that (if such exists) yields, when applied to $r$ and $s$, a common term.
One next applies that substitution (a most general unifier) to both $r$ and $e(s,t)$ to obtain two (so-to-speak) new hypotheses.
Finally, to the new pair of hypotheses, one applies detachment.
Clearly, any formula obtainable from an axiom set by condensed detachment alone can be obtained from that set by detachment together with substitution.
Conversely, it can be shown [5] that every formula obtainable with detachment and substitution is an {\em instance} of at least one formula obtainable with condensed detachment alone.

\subsection{History}
 
In 1933, J. {\L}ukasiewicz found the first three eleven-symbol formulas that could each serve as a single axiom for the equivalential calculus [6].
Expressed as clauses, they are the following.
(For ease of reference, we adopt the useful naming convention for formulas of this length devised later by Kalman and presented in the appendixes to [9] and [2].)
\begin{verbatim}
P(e(e(x,y),e(e(z,y),e(x,z)))). %1%
P(e(e(x,y),e(e(x,z),e(z,y)))). %1%
P(e(e(x,y),e(e(z,x),e(y,z)))). %1%
\end{verbatim}
In the same paper, {\L}ukasiewicz shows that no shorter formula can serve as a single axiom.
 
Meredith later discovered the following seven additional shortest single axioms [8].
\begin{verbatim}
P(e(e(e(x,y),z),e(y,e(z,x)))). %1%
P(e(x,e(e(y,e(x,z)),e(z,y)))). %1%
P(e(e(x,e(y,z)),e(z,e(x,y)))). %1%
P(e(e(x,y),e(z,e(e(y,z),x)))). %1%
P(e(e(x,y),e(z,e(e(z,y),x)))). %1%
P(e(e(e(x,e(y,z)),z),e(y,x))). %1%
P(e(e(e(x,e(y,z)),y),e(z,x))). %1%
\end{verbatim}
 
In the mid-1970s, J. Kalman and his student Peterson undertook a computer-assisted investigation of all 630 eleven-symbol equivalential theses (distinct up to alphabetical variance).
Kalman found another shortest single axiom among these [3], the eleventh.
(We note that Kalman himself graciously regards this eleventh axiom as simply correcting a misprint in [8].)
\begin{verbatim}
P(e(x,e(e(y,e(z,x)),e(z,y)))). %1%
\end{verbatim}
Peterson showed that 612 of the eleven-symbol theses were too weak [9], and posed as open questions the status of the remaining seven formulas of length eleven.
\begin{verbatim}
P(e(x,e(y,e(e(e(z,y),x),z)))). %1%
P(e(x,e(y,e(e(x,e(z,y)),z)))). %1%
P(e(x,e(e(e(e(y,z),x),z),y))). %1%
P(e(e(e(e(x,e(y,z)),z),y),x)). %1%
P(e(x,e(e(y,z),e(e(x,z),y)))). %1%
P(e(x,e(e(y,z),e(e(z,x),y)))). %1%
P(e(x,e(e(e(x,y),e(z,y)),z))). %1%
\end{verbatim}
 
Kalman's place in the history of equivalential calculus, and certainly in the history of this paper, is by no means limited to the preceding observations.
He was apparently the first to introduce automated reasoning and the equivalential calculus to each other, writing his own theorem-proving program to study possible shortest single axioms early on and using it to discover his proof for the eleventh shortest single axiom.
Later, having learned of the Argonne effort focusing on the automation of reasoning, he visited Argonne (in the late 1970s, if memory serves).
He brought a fine gift, the open questions concerning the status of those seven unclassified formulas of length eleven.
 
Not long after Kalman's visit, the Argonne group began an intense study of the seven unclassified formulas, conquering six of them.
It was proved that the first four of those formulas ($XJL$, $XKE$, $XAK$, and $BXO$) are too weak [12].
Of the remaining three formulas, S. Winker proved (as reported in [12]) that $XHN$ and $XHK$ are in fact new shortest single axioms, the twelfth and thirteenth.
\begin{verbatim}
P(e(x,e(e(y,z),e(e(z,x),y)))).%1%
P(e(x,e(e(y,z),e(e(x,z),y)))).%1%
\end{verbatim}
For the first of the two, Winker's proof has length 159 (applications of condensed detachment), for the second, length 83.
(We now have far shorter proofs, of lengths 19 and 21, respectively.)
 
One formula remained unclassified, as it had since 1977:  $XCB$.
With the goal of proving that this recalcitrant formula is in fact a single axiom, we sought to show that one could deduce from it either the 3-basis cited earlier or one of the thirteen previously known shortest single axioms.
(Indeed, because of the dependence of reflexivity, the independent 2-basis would have served well as a target; but, as a matter of historical fact, the 3-basis was one of the targets.)
 
Precisely how the mystery was solved to reveal $XCB$ as the fourteenth and final shortest single axiom is the focus of Section 2.
There we detail the methodology and strategy we used to obtain the original 71-step proof.
In Section 3, we discuss proof refinement and how we eventually obtained from the 71-step proof a far shorter proof, one of length 25.
There we present that 25-step proof, the shortest we know of.
The proof has been independently verified with two other programs.
In Section 4, we consider why $XCB$ remained unclassified for so many years and present additional observations on some unusual aspects of $XCB$.
Section 5 provides a summary.

\section{An Effective Methodology}
 
Here we present the highlights of our successful attack on the $XCB$ question.
In fact, a set of proofs showing that $XCB$ is a single axiom for the equivalential calculus was discovered, many of them completing with the deduction of one of the previously known shortest single axioms.
The shortest proof we found (given in Section 3) uses 25 applications of condensed detachment to reach,  rather than a single axiom, the independent 2-basis consisting of $e(e(x,y),e(e(y,z),e(x,z)))$ and $e(e(x,y),e(y,x))$.
Additional details concerning our approach, including some input files, summaries of output files, and useful commentary, are available on the Web page www.mcs.anl.gov/$\sim$wos/XCB/.
Perhaps one might have preferred to be shown a fully automated approach, but in fact no effective algorithm for answering such deep questions has yet been found.

In essence, the original attack involved three phases:  the deduction of reflexivity, the deduction of transitivity presuming the availability of symmetry, and the deduction of symmetry itself.
Of these three goals, the third proved by far the most difficult to reach.
Historically, the first and second goals were reached more than one year ago.
The recent effort was devoted to an attempt (successful in the end) to obtain a proof of symmetry from the formula $XCB$ alone.
The interim took the form of a long pause that witnessed essentially no research devoted to this intriguing formula.
Throughout, we were continually aware that reflexivity is easily proved to be dependent on symmetry together with transitivity.
Nevertheless, the 3-basis was originally one of our targets for proof completion.
Also included as targets were the thirteen previously known shortest single axioms for the equivalential calculus.
 
The attack relied on a number of strategies and features offered by OTTER.
Among these, the first chosen was {\em lemma adjunction}, which involves adjoining to $XCB$, in successive runs, the final steps of various intermediate targets proved in earlier runs or even the entire proofs of such targets.
The so-called lemmas that are adjoined are placed in the initial set of support, and the program is instructed to focus on each to initiate applications of the inference rule or rules in use before focusing on a newly deduced and retained conclusion.
These lemmas, rather than necessarily appearing as steps in the final sought-after proof, are intended to direct the program toward important steps that will appear in the final proof.
(The set of support strategy, forward subsumption, and usually back subsumption are almost always featured in our research with OTTER.)
For example, the inclusion initially of the eleven steps of our original proof of reflexivity as lemmas was not coupled with the expectation that any or all of those steps would appear in the final proof, if such were found.
In fact, as it turned out, its most complex step (of length 47, not counting the predicate symbol) does not appear in the proof we offer in Section 3, nor does it play much of a role throughout.
In contrast, the proof in Section 3 relies on two 47-symbol formulas, both of which occurred in a 52-step proof of $e(x,e(y,e(x,y)))$, completed in a breadth-first search, a proof obtained in approximately 78 CPU-hours.
The two formulas are, respectively, the 38th and 41st deduced steps of the 52-step proof.

In the search for a proof of symmetry from $XCB$, we chose as intermediate targets all fifteen of the 7-symbol theorems of the equivalential calculus with the 2-property.
(Hereafter, we shall suppress the phrase ``with the 2-property''.)
This choice was motivated in part by the fact that symmetry is such a theorem and in part because short formulas are, in our experience, often easier to prove than longer ones.
Of course, if all fifteen were proved, then symmetry (being such a 7-symbol theorem) would be proved, and the centerpiece of our work would be in hand.
 
For directing the program's reasoning, R. Veroff's {\em hints strategy} [11] played the prominent role.
To use that strategy, the researcher chooses one or more formulas or equations that the program treats as more important than other such items for initiating inference rule application.
In particular, when the program is ready to choose a new clause on which to focus, it prefers (if available) one that matches a hint.
We selected the option of choosing a matching clause or one that subsumes a hint.
We note that hints are used only to guide the program's reasoning; they are not used as hypotheses in the application of inference rules.
Rather, each is viewed as an attractive symbol pattern to be matched, if possible.
 
The hints that were used grew in number, each corresponding to a proof step of a target lemma proved in an earlier run.
That is, as the size of the initial set of support grew from run to run and experiment to experiment, so did the size of the list of hints.
To each hint, the researcher can assign a value; the smaller the value, the higher the priority for being chosen as the focus of attention.
By way of illustration, an included hint that corresponds to a formula of length 47 (measured in symbols) can in effect be treated as having length 1.
Indeed, various 47-symbol formulas occur in proofs of intermediate targets; among such formulas are the two found in our 25-step proof.
When the program is instructed to rely on complexity preference rather than on breadth first, the highest priority (for directing its reasoning) is given to formulas of least complexity.
The program computes complexity based on the assigned values to included resonators and hints, otherwise based on symbol count.
 
A second direction strategy was also employed, McCune's {\em ratio strategy} [14].
That strategy blends the choosing of focal clauses based on complexity (which, as indicated, can be influenced by included hints) and first-come first-serve.
If, for example, the value assigned to the pick\_given\_ratio is 2, then the program chooses two clauses by complexity, one by first-come first-serve, two, one, and so on.
Inclusion of the ratio strategy with a low assigned value to the pick\_given\_ratio permits the program to regularly focus on conclusions retained early, some of which may be very long and would otherwise be delayed in the context of inference rule initiation, perhaps forever.
 
To restrict the reasoning, the max\_weight parameter offered by OTTER proves most useful.
New conclusions whose complexity (that is, their weight, as determined by assigned values to resonators or hints or by symbol count) exceeds the value assigned to max\_weight are discarded.
A small assigned value can restrict the program's reasoning so severely that all proofs are blocked, while too large an assigned value can drown the program in newly deduced and retained information.
 
One additional restriction strategy was employed, at least in the early stages.
The strategy is a version of {\em term avoidance} (sometimes referred to as a {\em subtautology strategy}).
In early runs, we instructed the program to discard immediately any newly deduced conclusion that contains as a proper subformula a formula of the form $e(t,t)$ for some term $t$.
This action was taken because we feared that, otherwise, the space of conclusions to be explored would grow far too rapidly and prevent the program  from reaching the goal of a proof of symmetry.
The danger of its inclusion (which was in fact eventually experienced) is that key information on the path to a needed conclusion may be discarded, preventing the program from completing an assignment.
Later in the study, on the path that produced a proof of symmetry and more, the use of term avoidance was abandoned.
 
Finally, at the outermost level of directing the program's reasoning, we considered the choice between instructing the program to conduct its search by complexity preference (within the context of the ratio strategy) and by breadth first (that is, level saturation, in which the ratio strategy has no function).
Although both choices may require consideration of a space of conclusions that can grow exponentially, far more effective methodologies exist for coping with this possible growth with the first choice than with the second.
Nevertheless, the use of breadth first (level saturation) eventually did provide key results for both the lemma-adjunction phase and for the growing hints list.
We simultaneously and, as it turned out, profitably employed both of these global strategies.
 
Consistent with the plan of targeting the fourteen cited bases, the program commenced its attack.
In the spirit of lemma adjunction, we began (as noted) by relying on the year-old 11-step proof of reflexivity from $XCB$, using its steps both as lemmas and as hints.
(The single 47-symbol formula found among these eleven steps motivated us to instruct the program throughout our attack to retain formulas of this complexity.
The inclusion of the cited eleven lemmas enabled the program to probe far deeper levels than it would have been able to otherwise, providing us with additional proof steps to be used in later runs both as hints and as lemmas.)
We instructed the program to treat any formula that was identical to one of the eleven steps or that subsumed one of them as being of length 1.
Clearly, we were instructing the program to direct its reasoning with much preference for newly deduced clauses that matched or subsumed a hint.
(The hints strategy is generally preferred over the resonance strategy for studies in equivalential calculus because too many conclusions match a given resonator.)
 
OTTER's arsenal encourages attacks that rest on multiple approaches applied in separate but simultaneous runs.
We chose a two-pronged attack.
The sole difference between the two approaches that were chosen involved the means used to direct the program at the outermost level of its reasoning.
For one approach, we chose breadth-first search.
For the other, we chose McCune's ratio strategy, with an assigned value of 2 to the pick\_given\_ratio.
In both approaches, the value of 48 was assigned to the max\_weight.
In both, we also included nineteen hints corresponding to the steps of a proof we had in hand that the single axiom $XHN$ implies the single axiom $UM$.
This action was taken because we had abundant evidence of the difficulty of attempting to answer the open question concerning the axiomatic status of $XCB$ and because settling the question for $XHN$ had initially proved so difficult.
 
Both approaches began by employing the version of the term-avoidance strategy mentioned earlier.
Although the program could have been instructed to rely on the weighting strategy, instead we used demodulation to rewrite unwanted new conclusions to junk.
As noted, we were motivated by experience gleaned from much experimentation that teaches one that the space of deducible conclusions is sharply reduced with this strategy.
 
We placed in the passive list the negations of the intermediate targets, namely, the full set of fifteen 7-symbol theorems.
Also included were the negations of the known thirteen shortest single axioms, as well as the negation of transitivity.
Members of the passive list are used mainly to signal proof completion (by unit conflict) and for subsumption.
In general, we also use the passive list to monitor progress, viewing the proof of any of its members as a good sign even if we do not intend to use the result.
In this study, of course, the intention was to add as lemmas in a later run the deduced steps of all proofs of those intermediate 7-symbol targets.
 
In the input usable list (whose members never initiate inference rule application), we placed the clause that captures (with hyperresolution) condensed detachment.
\begin{verbatim}
-P(e(x,y)) | -P(x) | P(y).
\end{verbatim}
 
As noted, the key 7-symbol formula is symmetry, because our prior studies had shown that, if we had in hand a proof of symmetry deduced from $XCB$ alone, we could complete a proof that $XCB$ is indeed a shortest single axiom.
We of course had no way of knowing whether the final proof, if obtained, would include any other proved 7-symbol formulas or their proof steps.
Their adjunction was intended merely to aid the search in general, such adjunction of lemmas having proved to be a powerful methodology in many contexts in the past.
 
Another of the 7-symbol formulas, expressed in clausal form as P(e(e(e(x,y),x),y)), had been proved with some difficulty in an earlier experiment with a different program.
We strongly suspect that the discovery of that proof played the key role in our decision to include all 7-symbol theorems with the two-property as intermediate targets for eventual use in lemma adjunction.
The experiment in question would prove almost immediately to be valuable in evaluating the two-pronged approach, since it had not benefited from reliance on the strategies presented earlier.
Indeed, we were encouraged to continue pursuing both approaches because each obtained a proof of that clause.
The approach relying on the ratio strategy completed its proof in less than one CPU-minute; the breadth-first approach required approximately eleven CPU-minutes.
(In addition to any possible contributions from the 11-step proof of reflexivity, the length of the first proof is 13 and of the second is 17.
Since we were not seeking shorter proofs at that point, we note that a shorter proof may exist with $XCB$ as the sole hypothesis.)
Our success in proving the apparently difficult clause P(e(e(e(x,y),x),y)) offered a promise that would indeed be fulfilled.
 
The approach based on the use of the ratio strategy yielded no additional proofs and was temporarily discontinued.
The breadth-first approach, however, was permitted to continue its search (which, as will be seen shortly, was most fortunate).
While waiting for more proofs from the breadth-first search---if such could be found---the ratio-strategy approach was resumed, but with the addition (as lemmas) of the thirteen deduced steps of the proof just cited; our objective was to seek proofs of additional targets.
This effort failed.
 
An analysis suggested that, just perhaps, the use of the term-avoidance strategy might be blocking progress.
Therefore, we ceased using it but otherwise continued as in the preceding case.
The use of term avoidance was indeed the problem, at least temporarily.
Without it, four additional theorems of the fifteen were proved, making a total of six proved (because one of the fifteen is an instance of reflexivity), with nine yet to prove.
Of course, symmetry was still the only one of them crucial to the attack, the proofs of the others being of interest primarily for lemma adjunction and for supplying additional hints.
 
At this point, the decision to allow the breadth-first branch to continue brought riches.
Specifically, the first of the following two formulas was proved in approximately 45 CPU-hours and the second in approximately 78 CPU-hours.
\begin{verbatim}
P(e(e(x,e(y,x)),y)).
P(e(x,e(y,e(x,y)))).
\end{verbatim}
Completion of proofs for these two formulas signaled progress.
Of far greater importance, the steps in those proofs turned out to play a crucial role.
The join of the two proofs (of respective lengths 33 and 52) provided 71 additional formulas to use as lemmas for the next run.
 
In that run, finally, a proof of symmetry was completed.
An 18-step proof was found in approximately 45 CPU-minutes, a proof relying on nine of the proof-step lemmas that had been adjoined during the attack.
Deciding at this point not to rely on much earlier derivations, we next sought a proof of transitivity (where the target was the familiar 3-basis) or a proof of one of the known thirteen shortest single axioms.
Of course, our intention was to have OTTER prove directly from $XCB$ alone a known basis without reliance on any lemmas.
 
We chose another two-pronged approach.
On one branch, we relied on the far, far earlier set of hints that corresponded to a proof of transitivity from symmetry.
On the other branch, we ignored such earlier discoveries.
The second branch offered more appeal in that it corresponded more closely to a type of attack we enjoy.
For that approach, we relied on the hints used in the preceding run together with thirty hints corresponding to the join of the new proofs obtained in that run, among which was symmetry.
The only hypothesis that was used was $XCB$; the pick\_given\_ratio was assigned the value 2; the max\_weight was assigned the value 64, in case longer formulas might be useful; no term avoidance was employed.
McCune's program succeeded in finding (in approximately 7 CPU-seconds) a 61-step proof of symmetry and (in approximately 15 CPU-minutes) a 71-step proof of transitivity.
As expected, the latter does in fact depend on symmetry, its 66th step.
Our attack had vanquished $XCB$.
 
As for the first of the two approaches, it also succeeded and finished even earlier.
In approximately 4 CPU-seconds, a 66-step proof of transitivity was found, and in roughly 1 additional CPU-second symmetry was proved in 64 applications of condensed detachment.
The former, contrary to expectation, clearly does not depend on the latter.
In other words, aided by hints corresponding to results from more than one year earlier showing that the use of symmetry can lead to a proof of transitivity, a proof of the latter independent of symmetry was found.
Whereas the approach that relied upon hints proving transitivity from symmetry proved five of the previously known shortest single axioms, the other approach proved in approximately 45 CPU-minutes all thirteen.
 
Neither approach produced a complete proof of the 3-basis as a single proof, although each member of that basis was proved in each approach.
The type of proof we preferred would find a contradiction with the denial of the conjunction of the three members, thus providing a proof with no duplicate steps.
Our failure to find such a proof is explained by our failure to include a clause corresponding to the denial of the conjunction of the members of the 3-basis.
We simply reached the proof of the three members sooner than we had expected.
Based on our preferred approach, the type of proof we have in mind would be at least length 72 because reflexivity had not been proved in the 71-step proof, though symmetry had been proved.
(Had the denial of the 3-basis been included in the input, say, in the second approach, an experiment shows that indeed a 72-step proof would have been found.)
 
Therefore, with the goal of producing a proof of the 3-basis as a single proof rather than as three separate proofs (and with no duplicate steps), and with the additional goal of beginning a refinement study focusing on proof length, we turned to yet another run.
We simply took the input file for the second of the two approaches just described, added the denial of the 3-basis to the usable list, and added the command set(ancestor\_subsume).
(For those who are curious, the inclusion of this command ordinarily has a dramatic effect on CPU time, slowing the program sharply in many cases.)
That command instructs OTTER to compare derivation lengths with the same conclusion and prefer the shorter; it automatically seeks shorter proofs.
(We emphasize, however, that shorter subproofs do not necessarily a shorter total proof make.)
 
In fact, five proofs of the desired type were found of (in order) length 49, 46, 48, 47, and 42.
The first was completed in approximately 24 CPU-seconds and the fifth in approximately 15 CPU-minutes.
Remarkable to us and even startling, in approximately 14 CPU-hours the previously known thirteen shortest single axioms were each proved in less than 50 applications of condensed detachment.

\section{Proof and Refinement}
 
Had {\L}ukasiewicz, Meredith, or Prior, for example, been presented with the set of proofs that included the 71-step proof of transitivity, or better yet, the 42-step proof of the join that was found with ancestor subsumption, he would (we conjecture with virtual certainty) have embarked on a search for an abridgment, a shorter proof.
The discovery of shorter and simpler proofs was clearly also of interest to Hilbert and was the subject of his recently discovered twenty-fourth problem [10].
Long before learning of the Hilbert problem, much research by members of the Argonne group had been devoted to developing methodology for OTTER to apply in the context of proof refinement.
 
Naturally, therefore, next in order was the pursuit of a proof of length strictly less than 42 showing that $XCB$ is in fact a single axiom for the equivalential calculus.
Although no constraint was placed on the target---for example, any member of the thirteen previously known shortest single axioms would have been more than acceptable---we mainly continued to focus, perhaps for reasons of momentum, on the 3-basis.
Our approach was again iterative.
Each new and shorter proof found was, in general, used as a new target with its steps included as hints.
Throughout our attack, we continued to rely on ancestor subsumption.
In addition, we employed a technique called {\em demodulation blocking} to block the use of the steps of a given proof one at a time, with the objective of finding an abridgment.
This technique has proved effective in many contexts, whether the focus is on formulas or on equations, and in the presence of various inference rules.
 
When we reached a point at which neither ancestor subsumption nor demodulation blocking yielded an abridgment, we turned to the {\em cramming strategy} [15].
The object of (one incarnation of) the cramming strategy is to cram, or force, so many steps of a chosen subproof into new subproofs of the remaining members of a conjunction that the length of the new proof of the whole is strictly less than that which prompted the effort.
Intuitively, the object is to have the program focus on a subproof and have its steps do double duty, triple duty, or more.
For the target of the 3-basis, cramming was successfully used in two cases:  to cram on the subproof of transitivity and later (with a shorter proof of the 3-basis in hand) to cram on the subproof of symmetry.
 
During the refinement phase, OTTER eventually discovered a 26-step proof of the independent 2-basis.
OTTER also discovered a 27-step proof of the previously known single axiom $YQF$ and one of that length for the previously known single axiom $WN$.
With some thought, we were then able to shorten each of the three proofs by one step.
We next observed that, with an appropriate use of condensed detachment, the 25-step proof of the 2-basis could be extended to a 26-step proof of the 3-basis (including reflexivity).
 
We now present our 25-step proof of transitivity and then symmetry from $XCB$.
To aid one in reading OTTER's proofs, we included the command set(order\_history), a command that instructs the program to list the hypotheses (by number) of a deduced step in the order $i,j,k$, where $i$ is the clause for condensed detachment, $j$ the clause corresponding to the major premiss, and $k$ the clause corresponding to the minor premiss.
 
\begin{center}
{\bf A 25-Step Proof from $XCB$}
\end{center}
 
\begin{verbatim}
The command was "otter". The processID is 24362.

-----> EMPTY CLAUSE at   0.15 sec ---->
   134 [hyper,52,132,128] $ANSWER(all_s_t_indep).

Length of proof is 25.  Level of proof is 19.

---------------- PROOF ----------------

51 [] -P(e(x,y)) | -P(x) | P(y).
52 [] -P(e(e(a,b),e(b,a))) | -P(e(e(a,b),e(e(b,c),e(a,c)))) |
   ANSWER(all_s_t_indep).
53 [] P(e(x,e(e(e(x,y),e(z,y)),z))).
105 [hyper,51,53,53] P(e(e(e(e(x,e(e(e(x,y),e(z,y)),z)),
   u),e(v,u)),v)).
106 [hyper,51,105,53] P(e(e(e(e(x,e(e(e(x,y),e(z,y)),z)),
   u),v),e(u,v))).
107 [hyper,51,53,106] P(e(e(e(e(e(e(e(x,e(e(e(x,y),e(z,y)),
   z)),u),v),e(u,v)),w),e(v6,w)),v6)).
108 [hyper,51,106,53] P(e(x,e(e(e(e(e(y,e(e(e(y,z),e(u,z)),
   u)),x),v),e(w,v)),w))).
109 [hyper,51,107,53] P(e(e(e(e(e(e(e(x,e(e(e(x,y),e(z,y)),
   z)),u),v),e(u,v)),w),v6),e(w,v6))).
110 [hyper,51,106,108] P(e(x,e(e(e(e(e(y,e(e(e(y,z),e(u,z)),
   u)),e(e(v,e(e(e(v,w),e(v6,w)),v6)),x)),v7),e(v8,v7)),v8))).
111 [hyper,51,53,108] P(e(e(e(e(x,e(e(e(e(e(y,e(e(e(y,z),
   e(u,z)),u)),x),v),e(w,v)),w)),v6),e(v7,v6)),v7)).
112 [hyper,51,105,110] P(e(e(e(e(x,e(e(e(x,y),e(z,y)),z)),
   e(e(u,e(e(e(u,v),e(w,v)),w)),e(e(v6,e(e(e(v6,v7),
   e(v8,v7)),v8)),v9))),v10),e(v9,v10))).
113 [hyper,51,109,111] P(e(e(x,e(y,e(e(e(e(e(z,e(e(e(z,u),
   e(v,u)),v)),e(e(w,e(e(e(w,v6),e(v7,v6)),v7)),y)),v8),
   e(v9,v8)),v9))),x)).
114 [hyper,51,107,112] P(e(e(e(e(e(x,e(e(e(x,y),e(z,y)),
   z)),e(u,e(e(e(u,v),e(w,v)),w))),v6),v7),e(v6,v7))).
115 [hyper,51,53,113] P(e(e(e(e(e(x,e(y,e(e(e(e(e(z,
   e(e(e(z,u),e(v,u)),v)),e(e(w,e(e(e(w,v6),e(v7,v6)),
   v7)),y)),v8),e(v9,v8)),v9))),x),v10),e(v11,v10)),v11)).
116 [hyper,51,114,106] P(e(x,e(e(y,e(e(e(y,z),e(u,z)),
   u)),x))).
117 [hyper,51,53,116] P(e(e(e(e(x,e(e(y,e(e(e(y,z),
   e(u,z)),u)),x)),v),e(w,v)),w)).
118 [hyper,51,112,117] P(e(e(e(e(e(x,
   e(e(y,e(e(e(y,z),e(u,z)),u)),x)),v),w),e(v,w)),
   e(v6,e(e(e(v6,v7),e(v8,v7)),v8)))).
119 [hyper,51,112,118] P(e(e(e(e(e(e(x,e(e(y,e(e(e(y,z),
   e(u,z)),u)),x)),e(v,e(e(e(v,w),e(v6,w)),v6))),v7),
   v8),e(v7,v8)),e(v9,e(e(e(v9,v10),e(v11,v10)),v11)))).
120 [hyper,51,115,119] P(e(e(e(x,e(y,e(e(e(y,z),
   e(u,z)),u))),v),e(x,v))).
122 [hyper,51,120,105] P(e(e(e(x,e(e(e(x,y),e(z,y)),
   z)),e(e(e(u,v),e(w,v)),w)),u)).
123 [hyper,51,106,122] P(e(e(e(e(x,y),e(z,y)),z),x)).
124 [hyper,51,53,123] P(e(e(e(e(e(e(e(x,y),e(z,y)),
   z),x),u),e(v,u)),v)).
125 [hyper,51,124,123] P(e(e(e(x,y),x),y)).
127 [hyper,51,124,108] P(e(e(e(e(x,e(e(e(x,y),
   e(z,y)),z)),e(e(e(e(e(u,v),e(w,v)),w),u),
   v6)),v7),e(v6,v7))).
128 [hyper,51,127,123] P(e(e(x,y),e(e(y,z),e(x,z)))).
130 [hyper,51,128,125] P(e(e(x,y),e(e(e(z,x),z),y))).
131 [hyper,51,128,130] P(e(e(e(e(e(x,y),x),z),u),
   e(e(y,z),u))).
132 [hyper,51,131,123] P(e(e(x,y),e(y,x))).
\end{verbatim}
 
{\bf Open Question}.  Does there exist a proof of length 24 or less showing that $XCB$ is a single axiom, where the target is any of the other thirteen shortest single axioms for the equivalential calculus or is the 2-basis consisting of symmetry and transitivity?

\vspace{.1in}
With this proof in hand---but twenty-five steps in length---one might naturally wonder why it took so long to answer the $XCB$ question.
One answer rests in part with the fact that the iterative approach given in Section 2 was only recently formulated.
Perhaps a better answer rests with the behavior and, more important, the {\em apparent} behavior of $XCB$, which is the focus of next section.

\section{Close Inspection of $XCB$}
 
Insight, understanding, and ideas can sometimes be gained by asking why a question remained open for many years.
In this section, we consider various factors that may explain why intense effort and study failed to reveal the true status of $XCB$.
We also discuss some of the behavior and power of $XCB$, possibly providing insight that will aid in answering additional questions concerning axiomatizations of other formal systems.

\subsection{The Resistance of the $XCB$ Question}
 
Our study of $XCB$ suggests several possible factors that may have contributed to its resistance to classification.
One such factor concerns a property that one might easily guess (though incorrectly, as it turns out) is shared by all of the theorems deducible from $XCB$.
Indeed, the theorems that are readily deducible from $XCB$ using our usual strategies contain---as do the first seventeen steps of the proof presented in Section 3---at least one alphabetical variant of $XCB$ as a subformula.
Therefore, it is not surprising that those researchers credited with the dispatching of $XJL$ and its three equally weak cousins asserted that $XCB$ was also too weak, that all theorems deducible from it must contain a variant of XCB.
(That claim, made in [12], was corrected in [14].)
In fact, as some of our later experiments using breadth-first search show, all of the 1,494 theorems obtainable from $XCB$ through the first five levels contain such an alphabetic variant.
Moreover, with the single exception of the formula $e(e(x,y)$,$e(x,y))$---which of course refutes the conjecture---so do the additional 319,493 theorems obtainable at level six.
 
Had the conjecture held, the theorem $e(x,x)$ would have been out of reach (not deducible) with $XCB$ as hypothesis and condensed detachment as the sole rule of inference.
Instead, a simple experiment relying on a breadth-first search and consideration of formulas of complexity less than or equal to 35 (not counting the predicate symbol) shows that reflexivity is in fact provable from $XCB$ with but eleven applications of condensed detachment, a proof obviously different from the one found more than a year ago (which contained a 47-symbol formula).
Further, a limit on complexity of 31 even suffices, producing a 17-step proof.
This simple formula eluded capture, however, in part because (some of) the authors and their colleagues overlooked for far too long the possible value of such a search and in part because of their conjecture that reflexivity was not provable.
Among the truths about research are:  The knowledge that a result holds seems to make its proof easier to complete; the belief that it does not disposes one to follow the wrong path if that belief is mistaken.
 
The discovery of a proof of reflexivity was encouraging.
But, even after this discovery, $XCB$ continued to resist classification.
Other factors were contributing to its resistance.
For an example of one such factor, recall that condensed detachment proceeds by unifying the antecedent (leftmost major argument) of the major premiss with the minor premiss.
Common to the majority of proofs we have discovered is the application of condensed detachment to the following two formulas (expressed as clauses) of complexity 47, with the first as major and the second as minor premiss.
\begin{verbatim}
P(e(e(e(e(e(x,e(y,e(e(e(e(e(z,e(e(e(z,u),e(v,u)),v)),
   e(e(w,e(e(e(w,v6),e(v7,v6)),v7)),y)),v8),e(v9,v8)),
   v9))),x),v10),e(v11,v10)),v11)). %1%
P(e(e(e(e(e(e(x,e(e(y,e(e(e(y,z),e(u,z)),u)),x)),
   e(v,e(e(e(v,w),e(v6,w)),v6))),v7),v8),e(v7,v8)),
   e(v9,e(e(e(v9,v10),e(v11,v10)),v11)))). %1%
\end{verbatim}
 
One can easily imagine how daunting would be the prospect of applying by hand condensed detachment to this pair of formulas.
As measured in symbol count, the most general common instance of the antecedent of clause 72 and of all of clause 76 has complexity 2919.
 
The first completed (42-step) proof of the 3-basis contained these two formulas and one other of the same complexity (47).
We know of no proof that is free of 47-symbol formulas, although we do have in hand a proof with but one 47-symbol formula, different from the two just displayed.
By comparison, a proof for $XHN$---the last of the previously known single axioms to be found---can be completed with complexity limited to 35.

\subsection{The Behavior of $XCB$}
 
The behavior of the formula $XCB$ is dramatically different from that of the other shortest single axioms in a number of ways.
For one difference, with $XCB$ as the sole hypothesis, the range of the lengths of the proofs of the remaining thirteen axioms is rather small, from 26 to 30 steps.
Our experiences with other areas of logic, with other areas of mathematics, and (most important) with the other shortest single axioms had never before revealed such clustering of proof lengths.
The fourteen single axioms can thus be arranged so that $XCB$ is at the center of a ring with the other thirteen shortest single axioms roughly equidistant from it.
In contrast, the shortest path lengths we have found from the single axiom $UM$ to the other shortest single axioms vary sharply.
In particular, $XGF$ follows in a single application of condensed detachment, while, at the other end of the spectrum, $XCB$ requires twenty-three.
 
To test whether this is the case for other shortest single axioms, we focused on $XHN$.
We used as hints the 20 steps of our proof that $UM$ can be deduced from $XHN$.
Simultaneously, with $XCB$ as the hypothesis, we used as hints the 26 steps of our proof of the 3-basis.
Each experiment ran for more than twenty-three CPU-hours on a 400 MHz computer.
When $XCB$ was the sole hypothesis, the proofs of the other thirteen shortest single axioms ranged in length (as noted) from 26 to 30.
In contrast, when $XHN$ was the sole hypothesis, the proof lengths of the other thirteen single axioms ranged from 19 to 37.
In the two experiments under discussion, ancestor subsumption was used with the goal of finding ``short'' proofs.
We remark that the length of a proof in no way suggests how deep is the theorem that is proved nor how hard it is to find a proof.
Perhaps a better indication of difficulty is offered by the number of years a question remains open:  less than four years sufficed for dispatching $XHN$, twenty-five years for $XCB$.
 
For $XCB$, the proofs we have found so far require twelve distinct variables, whereas nine suffice for seeking proofs relying on $XHN$.
Whether a proof for $XCB$ exists relying on strictly fewer than twelve distinct variables is not yet known.
In contrast, we strongly conjecture that for $XHN$ the lower bound is nine.
 
We conclude this section by noting that we have also discovered two intriguing 27-step proofs relying solely on $XCB$ (and, of course, condensed detachment), the first completing with $YQF$ and the second with $WN$, each a shortest single axiom for equivalential calculus.
The two proofs agree on their first twenty-six steps.
Moreover, the last step of each of the two proofs is obtained by applying condensed detachment to the same pair of formulas, which indeed implies that the role of major and minor premiss is exchanged.
This striking phenomenon was certainly new to us.

\section{Summary}
 
In logic and in mathematics, once axioms have been found for an area of interest, a natural question concerns the existence of a single axiom.
If such is found, one might then seek to find a ``short'' single axiom.
If success again occurs, next comes the question of the existence of one or more shortest single axioms.
 
With the main result of this article in hand, the set of answers to those questions about equivalential calculus is complete.
There indeed do exist single axioms for that calculus.
The shortest have length eleven (measured in symbols), and exactly fourteen such axioms exist.
Thus the search for shortest single axioms for the equivalential calculus is at an end.
 
Perhaps the story of these formulas unfolded in an appropriate way:  $XCB$, the fourteenth and last of the shortest single axioms to be found, is unique among the fourteen in various ways and appears to have been the most difficult to study.

\section*{References}
 
\vspace{.1in}
\noindent
1.  Belnap, N., ``The two-property'', {\em Relevance Logic Newsletter} {\bf 1} (1976) 173--180.
 
\vspace{.1in}
\noindent
2.  Hodgson, K., ``Shortest single axioms for the equivalential calculus with CD and RCD'', {\em J. Automated Reasoning} {\bf 20} (1998) 283--316.

\vspace{.1in}
\noindent
3.  Kalman, J. A., ``A shortest single axiom for the classical equivalential calculus'', {\em Notre Dame J. Formal Logic} {\bf 19} (1978) 141--144.
 
\vspace{.1in}
\noindent
4.  Kalman, J. A., ``The two-property and condensed detachment'', {\em Studia Logica} {\bf 41} (1982), 173--179.
 
\vspace{.1in}
\noindent
5.  Kalman, J. A., ``Condensed detachment as a rule of inference'', {\em Studia Logica} {\bf 42} (1983), 443--451.
 
\vspace{.1in}
\noindent
6.  {\L}ukasiewicz, J., {\em Selected Works,} edited by L. Borokowski, North Holland, Amsterdam, 1970.
 
\vspace{.1in}
\noindent
7.  McCune, W., {\em OTTER 3.0 Reference Manual and Guide,} Tech. Report ANL-94/6, Argonne National Laboratory, Argonne, Illinois, 1994.
 
\vspace{.1in}
\noindent
8.  Meredith, C. A., and Prior, A., ``Notes on the axiomatics of the propositional calculus'', {\em Notre Dame J. Formal Logic} {\bf 4}, no. 3 (1963) 171--187.
 
\vspace{.1in}
\noindent
9.  Peterson, J. G., {\em The Possible Shortest Single Axioms for EC-Tautologies}, Report 105, Department of Mathematics, University of Auckland, 1977.
 
\vspace{.1in}
\noindent
10.  Thiele, R., and Wos, L., ``Hilbert's twenty-fourth problem'', {\em J. Automated Reasoning}, to appear.
 
\vspace{.1in}
\noindent
11.  Veroff, R., ``Using hints to increase the effectiveness of an automated reasoning program: Case studies'', {\em J. Automated Reasoning} {\bf 16}, no. 3 (1996) 223--239.
 
\vspace{.1in}
\noindent
12.  Wos, L., Winker, S., Veroff, R., Smith, B., and Henschen, L., ``Questions concerning possible shortest single axioms for the equivalential calculus: An application of automated theorem proving to infinite domains'', {\em Notre Dame J. Formal Logic} {\bf 24} (1983) 205--223.
 
\vspace{.1in}
\noindent
13.  Wos, L., ``Meeting the challenge of fifty years of logic'', {\em J. Automated Reasoning} {\bf 6} (1990) 213--232.
 
\vspace{.1in}
\noindent
14.  Wos, L., and Pieper, G. W., {\em A Fascinating Country in the World of Computing: Your Guide to Automated Reasoning,} World Scientific, Singapore, 1999.
 
\vspace{.1in}
\noindent
15.  Wos, L., ``The strategy of cramming'', Preprint ANL/MCS-P898-0801, Mathematics and  Computer Science Division, Argonne National Laboratory, Argonne, Illinois, 2002.
 
\vspace{.1in}
\noindent
16.  Wos, L., Ulrich, D., and Fitelson, B., ``XCB, the last of the shortest single axioms for the classical equivalential calculus'', {\em Bulletin of the Section on Logic}, accepted.
 
\end{article}
\end{document}